\documentclass{bmcart}

\usepackage[version=3]{mhchem} 

\usepackage{natbib}
\usepackage[utf8]{inputenc} 

\usepackage{graphicx} 

\startlocaldefs
\endlocaldefs

\begin{document}

\begin{frontmatter}

\begin{fmbox}
\dochead{Research}


\title{Physical and Electrochemical Area Determination of Electrodeposited Ni, Co, and NiCo Thin Films}


\author[
   addressref={aff1,aff2}, 
]{\inits{MJ}\fnm{Matthew J} \snm{Gira}}
\author[
   addressref={aff1},
   noteref={n1},
]{\inits{KP}\fnm{Kevin P} \snm{Tkacz}}
\author[
   addressref={aff1},
   corref={aff1},
   email={hampton@hope.edu}
]{\inits{JR}\fnm{Jennifer R} \snm{Hampton}}


\address[id=aff1]{
  \orgname{Department of Physics, Hope College}, 
  \city{Holland, MI 49423}, 
  \cny{USA} 
}
\address[id=aff2]{
  \orgname{Department of Chemistry, Hope College}, 
  \city{Holland, MI 49423}, 
  \cny{USA} 
}


\begin{artnotes}
\note[id=n1]{\emph{Present address:} Department of Chemical Engineering and Materials Science, University of California, Irvine, Irvine, CA 92697, USA} 
\end{artnotes}



\begin{abstractbox}

\begin{abstract} 
The surface area of electrodeposited thin films of Ni, Co, and NiCo was evaluated using electrochemical double-layer capacitance, electrochemical area measurements using the \ce{[Ru(NH3)6]^{3+}}/\ce{[Ru(NH3)6]^{2+}} redox couple, and topographic atomic force microscopy (AFM) imaging. These three methods were compared to each other for each composition separately and for the entire set of samples regardless of composition. Double-layer capacitance measurements were found to be positively correlated to the roughness factors determined by AFM topography. Electrochemical area measurements were found to be less correlated with measured roughness factors as well as applicable only to two of the three compositions studied. The results indicate that \emph{in situ} double-layer capacitance measurements are a practical, versatile technique for estimating the accessible surface area of a metal sample.
\end{abstract}



\end{abstractbox}
\end{fmbox}

\end{frontmatter}




\section*{Background}

Nanoporous materials are of increasing scientific and technological interest due to a variety of useful properties such as low mass density, high surface area, high strength, and enhanced optical, electrical, thermal, and catalytic behavior. Potential applications of metals with nanoporous morphology include batteries, capacitors, magnetic storage media, lightweight structures, sensors, and water filtration devices~\cite{tappan_2010}. The enhanced surface area and size-dependent reactivity of nanoporous metals also make them a promising area of study for a number of catalytic applications.

An important factor in evaluating the reactivity of a porous metal is the surface area available for reaction. Both increased surface area and changes in intrinsic reactivity can have significant effects on the overall behavior of a target material. Thus, straightforward and practical area measurement procedures are an essential aspect of catalysis research.

One technique for area measurement is based on the physical absorption of gas molecules to a surface following the theory presented by Brunauer, Emmet, and Teller (BET)~\cite{brunauer_1938, ji_2003, snyder_2008}. Although this is a well-understood and regularly-used method, BET measurements have limitations, specifically the effects that heat treatments may have on the sample being characterized as well as the larger sample sizes needed to achieve the desired sensitivity~\cite{liu_2009}.

Electrochemical techniques for determining surface area have the advantage of being \emph{in situ} and can be performed just previous to or after any electrochemical reactivity measurements of interest. These techniques fall into two general categories. The first type uses a surface-limited chemical reaction to quantify the surface area of the electrode. In contrast, the second type measures a physical characteristic that is proportional to the surface area.

Using a surface-limited chemical reaction such as adsorption of hydrogen or carbon monoxide~\cite{correia_1997, liu_1997, lakshminarayanan_1997, shervedani_1999, de_souza_2000, sogaard_2001, vidakovic_2007, fang_2010, lindstrom_2010, schulenburg_2010}, underpotential deposition of a new metallic species~\cite{creus_1992, liu_2009, fang_2010, rouya_2012}, or surface oxide formation~\cite{machado_1994, jarzabek_1997, shervedani_1999, lukaszewski_2010, rouya_2012, grden_2012, van_drunen_2013} to quantify the surface area of the electrode can be quite sensitive. However, a disadvantage is that a particular reaction may be specific to the material being assessed. For example, gold oxide formation has been used extensively as a probe of gold electrode surface area, but this method can not be applied directly to an electrode of a different composition without considering the extent and potential range of oxide formation on that new material.

Rather than a chemical reaction, a electrochemical characterization using a physical characteristic can be used to quantify the surface area of a working electrode. The current due to a well-characterized redox reaction, such as the reduction of \ce{[Fe(CN)6]^{3-}} to \ce{Fe(CN)6]^{4-}}, is one such measurement~\cite{jarzabek_1997, gregoire_2009, wozniak_2010, patel_2012}. Similarly, the electrochemical double-layer capacitance of an electrode, which can be measured either by cyclic voltammetry or by electrochemical impedance spectroscopy, is proportional to its surface area~\cite{zoltowski_1993, shervedani_1999, simpraga_1995, jarzabek_1997, simpraga_1997, simpraga_1998, kellenberger_2005, lukaszewski_2010, tremblay_2010, herraiz-cardona_2011, grden_2012, patel_2012, hasegawa_2012, van_drunen_2013}. These techniques depend on the conducting nature of the electrode rather than its chemical identity, so to first approximation they do not depend on the nature of the material being studied. However, the potential range necessary for these measurements must be considered, because the characterization technique itself may affect the structure or composition of the material in question.

Topographic measurements of samples with a scanning tunneling microscope (STM) or atomic force microscope (AFM) can also be used to quantify the surface area of a sample~\cite{lakshminarayanan_1997, lust_1998, saffarian_1998}. These methods have the advantage of providing direct quantitative measurements of surface morphology. For AFM in particular, topographic measurements are not sensitive to the nature of the surface being probed. However, scanning probe techniques are local rather than ensemble measurements. Thus, a number of images must be taken for any surface in question to ensure the images are representative of the sample as a whole. For materials with porous morphology, scanning probe microscope measurements are limited, because the local probe can only measure structures which are accessible from the top of the sample. Similarly, if a surface has features smaller than that of the scanning probe tip itself, those features will not be imaged accurately by the technique. However, for materials with simpler morphology, scanning probe measurements provide a nice complement to the other methods described here.

In this work we compare electrochemical methods for determining the surface area of electrodeposited metal thin films with AFM topographic measurements of the same samples. Electrodeposited nickel, cobalt, and nickel-cobalt were chosen for the study because of the interest in these materials as catalysts. The thickness of these films was varied by controlling the total charge during the deposition process. In this way, the resulting roughness, and therefore surface area, of the material was varied. The resulting films were characterized using two electrochemical methods, double-layer capacitance measurements and area determination using a ruthenium-based redox probe. These measurements were compared to the roughness factors extracted from \emph{ex situ} AFM images of the samples. Correlations between these three measurements were explored, both for the samples with the same composition and for the entire set of samples regardless of composition.

\section*{Methods}

\subsection*{Electrochemistry}

The electrodeposition and electrochemical characterization were performed using an Epsilon electrochemical workstation (Bioanalytical Systems, Inc., West Lafayette, IN, USA) and a custom-built Teflon cell with a working electrode area of 0.032~cm$^2$ defined with a Kalrez o-ring~\cite{wozniak_2010}. The counter electrode was a coil of platinum wire (Alfa Aesar, Ward Hill, MA, USA) and the reference electrode was an \ce{Ag}/\ce{AgCl} (3~M \ce{NaCl}) electrode (Bioanalytical Systems, Inc., West Lafayette, IN, USA). All of the potentials recorded are with respect to this reference electrode. The electrolyte solutions were created using water that was purified through successive reverse osmosis, deionization, and UV purification stages. All of the chemicals used for these electrolytes were purchased from Sigma-Aldrich (St. Louis, MO, USA) and used as received. Every experiment was carried out at room temperature.

\subsection*{Deposition}

All thin films were deposited from solutions containing 0.5~M \ce{H3BO3} and 1~M \ce{Na2SO4} along with 0.1~M \ce{NiSO4} for the nickel thin films, 0.1~M \ce{CoSO4} for the cobalt thin films, or 0.75~mM \ce{NiSO4} and 0.25~mM \ce{CoSO4} for the nickel-cobalt thin films. The working electrode substrates were cleaved from a silicon wafer plated with 1000~\AA\ of gold over a 50~\AA\ titanium adhesion layer (Platypus Technologies, LLC, Madison, WI, USA). Controlled potential electrolysis was used to step the potential of the working electrode from open circuit to $-1000$~mV. The deposition was stopped once the desired amount of charge, ranging from 200~mC to 1000~mC, was achieved in order to vary the thickness of the deposited films.

\subsection*{Physical Characterization}

Physical characterization of the samples consisted of roughness and composition measurements. Atomic force microscope topography was used to measure the roughness of each thin film. This was completed using a Dimension Icon AFM (Bruker, Santa Barbara, CA, USA) using the ScanAsyst mode and SCANASYST-AIR cantilevers. A minimum of three 10~$\mu$m AFM images (512 pixels $\times$ 512 pixels) were taken of each sample. Nanoscope Analysis software (Bruker, Santa Barbara, CA, USA) was used to find the three-dimensional area of each image. For the NiCo thin films the elemental composition was measured. Scanning electron microscopy (SEM) and energy dispersive x-ray spectroscopy (EDS) measurements were completed using a TM3000 Tabletop SEM (Hitachi, Tokyo, Japan) and a Quantax 70 EDS attachment (Bruker, Madison, WI, USA). Images and EDS data were taken at $\times$60 magnification, and Quantax 70 software was used to obtain the Ni and Co compositions from the EDS spectra.

\subsection*{Electrochemical Characterization}

Electrochemical characterization consisted of double-layer capacitance and active area measurements. Electrochemical capacitance was measured using cyclic voltammetry (CV) in 0.5~M \ce{KOH} by sweeping from $-50$~mV to $-350$~mV and back to $-50$~mV. The scan rates were varied between 25~mV/s and 400~mV/s. The electrochemically active area was also measured using CV. The electrolyte solution was 5~mM \ce{Ru(NH3)6Cl3} and 1~M \ce{KCl}. The potential was swept from 100~mV to $-600$~mV and back to 100~mV with varying scan rates in the range of 100~mV/s to 901~mV/s. A minimum of three trials of both experiments were performed for each sample.

\section*{Results and Discussion}

The goals of this work were to explore the correlations between the AFM-based and electrochemical measurements for samples with different roughnesses and therefore different areas. The roughness of each of the samples was determined using AFM topographic measurements. Example AFM images are shown in Figure~\ref{fig:AFMEx} for samples with a deposited charge of 1000~mC. The Ni and Co films exhibit similar crystallite formation, with the resulting Co features a larger and taller than the corresponding Ni ones for the same deposited charge. In contrast, the NiCo film has a distinct texture with smaller, less compact crystallites.

For each image, the data were flattened using a first order filter to remove sample tilt. Afterwards, the roughness factor, $RF$, was calculated as $RF = A_\mathrm{AFM}/A_\mathrm{proj}$, where $A_\mathrm{AFM}$ is the surface area calculated from the image using the Nanoscope Analysis software and $A_\mathrm{proj}$ is the projected (flat) area of the measured region, 100~$\mu$m$^2$ in this case. From this calculation, the roughness factor is proportional to the surface area of the sample measured using AFM, but is not specific to the image sized used.

The average $RF$ for the three types of films are graphed in Figure~\ref{fig:RFvQ} as a function of the deposited charge, $Q$. The approximate average thickness, $t$, of the films corresponding to each deposited charge is shown on the upper horizontal axis of the figure. The conversion from deposited charge to thickness was calculated assuming 100\% current efficiency from $t = Q/(neA\rho^*)$, where $n = 2$ is the number of electrons in the Ni or Co deposition reaction, $e$ is the charge on the electron, $A$ is the defined area of the working electrode, and $\rho^*$ is the number density of the deposit. The bulk densities (in g/cm$^3$) and molar masses (in g/mol) of Ni and Co were used to calculate a value of $\rho^*$ for each metal. Because the values for Ni and Co are so similar, $9.14 \times 10^{22}$~cm$^{-3}$ and $9.09 \times 10^{22}$~cm$^{-3}$ respectively, an average value of $\rho^*$ was used to calculate the axis in the figure, corresponding to the assumption of an equal-component alloy. The systematic error for this assumption compared to using the value of $\rho^*$ for pure Ni or pure Co is approximately 0.2\%.

As seen qualitatively in Figure~\ref{fig:RFvQ}, for the same film thickness, the Ni samples generally are the smoothest, the Co samples have the roughest topography, and the NiCo alloy samples have intermediate roughness factors. For the Ni and Co samples, the roughness factor generally increases as the thickness of the samples increases, while for the NiCo samples, the roughness fluctuates with deposited charge. For the entire set of samples, regardless of composition, the roughness factors ranged from about 1.05 to 1.4. That is the samples had measured surface areas ranging from 5\% to 40\% higher than the corresponding projected area.

The compositions of the NiCo thin films were measured from EDS spectra taken at $\times$60 magnification and are shown in Figure~\ref{fig:NivsRF} as a function of the average roughness of the samples. The Ni composition of the deposited alloys was generally between 60 and 70~at. \%. The fact that the samples have a smaller Ni composition than that of the deposition solution (75~at. \%) is attributed to the anomalous codeposition phenomenon which is common for iron group metals~\cite{brenner_1963, akiyama_1992, sasaki_1998, zech_1999}. The decrease in Ni composition with increasing roughness is consistent with the data in Figure~\ref{fig:RFvQ} where NiCo samples are generally rougher than the Ni samples but smoother than the Co samples.

Electrochemical double-layer capacitance measurements were made on all the electrodeposited samples using CV in \ce{KOH} electrolyte. Example measurements for a variety of scan rates are shown in the inset of Figure~\ref{fig:CapEx}, showing the featureless current response expected of a capacitor. For these metals in alkaline electrolytes, a more complex pseudocapactive response corresponding to metal oxide and/or hydroxide redox reactions is often seen~\cite{hu_2002, chang_2010, kong_2011}. For the measurements here, however, the potential window used is significantly negative of that needed for these redox reactions to occur. As a result, the featureless CVs shown in the inset are measured instead. For a given scan rate, $v$, the average currents during the forward and reverse sweeps were calculated, and half of the difference between these two values was taken as the capacitive current, $I_\mathrm{dl}$, for that scan rate. This current was linearly dependent on the scan rate, as seen in Figure~\ref{fig:CapEx}, indicating that the films acted as simple capacitors in this potential scan range. The measured capacitance, $C_\mathrm{dl}$, was calculated using the time derivative of the definition of capacitance, $I_\mathrm{dl} = C_\mathrm{dl}v$, as the slope of the linear fit~\cite{bard_2001}.

Electrochemical area measurements were made on Ni and NiCo samples with CV using the \ce{[Ru(NH3)6]^{3+}}/\ce{[Ru(NH3)6]^{2+}} redox couple. Example measurements for a variety of scan rates are shown in the inset of Figure~\ref{fig:AreaEx}, which show the expected current response for a reversible redox reaction~\cite{bard_2001}. The ruthenium-based probe was chosen because the potential window for the CV experiment generally does not interfere with the deposited film. These area measurements could not be made on the Co samples, however, because the CV measurements did not result in reversible redox behavior and the scans in that potential range affected the structure of the film. The magnitude of the peak cathodic current, $I_\mathrm{p}$, as a function of the scan rate, $v$, is shown in Figure~\ref{fig:AreaEx} for an example measurement. The electrochemical area, $A_\mathrm{ec}$, of the sample was calculated using the Randles-Sevcik equation, $I_\mathrm{p} = 0.4463 n F A_\mathrm{ec} C (nF/RT)^{1/2} v^{1/2} D^{1/2}$, where $n = 1$ is the number of electrons involved in the redox reaction, $F$ is Faraday’s constant, $C$ is the bulk concentration of the analyte, $R$ is the molar gas constant, $T$ is the temperature, and $D$ is the diffusion constant of the analyte~\cite{bard_2001}. For \ce{[Ru(NH3)6]^{3+}}, the measured diffusion constant is $7.1 \times 10^{-6}$~cm$^2$/s~\cite{bard_1986, licht_1990, engblom_2001}.

The results of these two electrochemical measurements, the average $C_\mathrm{dl}$  and $A_\mathrm{ec}$ for each sample, are graphed as a function of the AFM-based measurement results, average $RF$, in Figure~\ref{fig:CAvsRF}(a) and (b) respectively. Because $A_\mathrm{ec}$ could not be measured for the Co samples, no data for Co are included in Figure~\ref{fig:CAvsRF}(b).

The results in Figure~\ref{fig:CAvsRF}(a) for all three types of samples show that there is a clear trend towards larger capacitance for rougher samples. There is some fluctuation in this correlation between capacitance and roughness, which increases for the rougher samples. Within this level of fluctuation, however, the observed trend between capacitance and roughness factor is the same for the group of samples as a whole, regardless of the sample composition or the morphological differences seen in the AFM topography (Figure~\ref{fig:AFMEx}). This was of particular interest for this study because of the practical importance of determining surface area of materials with a variety of compositions and structures. For these reasons, the results indicate that electrochemical double-layer capacitance is useful as a semi-quantitative measure of the surface area of electrodeposited samples.

In contrast to the capacitance results, the correlation between area measurements and roughness factor, shown in Figure~\ref{fig:CAvsRF}(b) for the Ni and NiCo samples, is less clear. In particular, although the smoother Ni samples generally have lower capacitance values than the rougher NiCo samples, they have higher measured electrochemical areas.

To explore these observations further, the ratio of average capacitance to average area, $C_\mathrm{dl}/A_\mathrm{ec}$, was calculated for each of the Ni and NiCo samples. Figure~\ref{fig:CAvsRF}(c) shows this ratio as a function of the average $RF$ of the samples. For the Ni samples, the capacitance-to-area ratio fluctuates between 40 and 75~$\mu$F/cm$^2$ for all roughness factors. This value is larger than, but on the order of 20 $\mu$F/cm$^2$, the specific capacitance value typically used in the literature for a variety of metals and alloys~\cite{trasatti_1991, chen_1991, chen_1992, zoltowski_1993, shervedani_1997a, simpraga_1997, shervedani_1997b, shervedani_1998, shervedani_1999, kellenberger_2005, tremblay_2010, fang_2010, herraiz-cardona_2011, grden_2012, van_drunen_2013}. In contrast, the NiCo films have even larger capacitance-to-area ratios, between 100 and 500~$\mu$F/cm$^2$, and the ratio tends to increase with increasing roughness factor. The larger ratios for the NiCo films may be the result of the area measurements being smaller than they should be. Additional evidence for this interpretation is seen in Figure~\ref{fig:CAvsRF}(b), where the NiCo area measurements are generally smaller than the Ni area measurements of samples with similar roughness factors.

One explanation for the electrochemical areas of the NiCo samples being underestimated is that in addition to the NiCo films generally being rougher than the Ni films, they display a distinct morphology (Figure~\ref{fig:AFMEx}(c)). For rougher, more complex morphologies, the assumption of planar diffusion which leads to the Randles-Sevcik equation may not be accurate. Specifically, the thickness of the diffusion layer can be as large as $10$s of $\mu$m for the scan ranges and rates used in the area measurements~\cite{bard_2001}. Thus, for the samples here, with topographic features on the scale of $100$s of nm to a few $\mu$m, some portions of the sample area would not contribute as strongly to the measured current compared to that expected from the simple planar diffusion model. On the other hand, double-layer capacitance measurements do not depend on the geometry and extent of the diffusion layer. Instead, during capacitive charging and discharging, non-specifically adsorbing ions such as \ce{K$^+$} and \ce{OH$^-$} can approach an electrode surface as close as the outer Helmholtz plane, generally a distance of 5 to 10 \AA~\cite{bard_2001}. Thus, area measurements may be underestimated in the case of rough, complex topography compared to capacitance measurements of the same sample. This, in turn, would lead to the observed higher capacitance-to-area ratios as well as to the lack of correlation between area and roughness measurements. A similar, but smaller, effect may also explain capacitance-to-area ratios for the smoother Ni samples being slightly higher than is typical in the literature.

\section*{Conclusions}

For the metal thin films studied here, the results indicate that \emph{in situ} electrochemical measurements of double-layer capacitance are correlated with the roughness factors extracted from \emph{ex situ} topographic AFM images. In addition, these measurements can be adapted to a wide variety of metal systems by choosing an appropriate potential range where only capacitive behavior is evident, thus minimizing any permanent effects on the sample. In contrast, the area measurements using a ruthenium-based redox probe are both less correlated with roughness measurements and less broadly applicable.

The fluctuations present in the capacitance vs.\ roughness data do place some limitations on the quantitative nature of the results. Nevertheless, the versatility and simplicity of capacitance measurements make the technique useful as a semi-quantitative measure of the electrochemically accessible surface area of a sample. Ongoing work in our lab aims to explore this method further by looking at additional metals and alloys as well as at the more complex morphologies with higher roughness factors, such as those produced by electrodeposition through self-assembled colloidal sphere masks. Double-layer capacitance provides a simple, practical, and reliable measure of the accessible surface area of metal and alloy thin films which can be used to quantify the intrinsic reactivity of these systems towards a variety of catalytic reactions.


\begin{backmatter}

\section*{Competing interests}
The authors declare that they have no competing interests.

\section*{Author's contributions}
MJG and KPT carried out the experiments and contributed to the data analysis. JRH coordinated the study and helped analyze the data. All authors helped to draft the manuscript and approved its final form.

\section*{Acknowledgements}
This material is based upon work supported by the United States National Science Foundation under grants no. RUI-DMR-1104725, REU-PHY/DMR-1004811, MRI-CHE-1126462, MRI-CHE-0959282, and ARI-PHY-0963317 as well as by the Jacob E. Nyenhuis Faculty Development Fund of Hope College.


\bibliographystyle{bmc-mathphys} 
\bibliography{afm-rough} 




\section*{Figures}
\begin{figure}[h!]
    \includegraphics{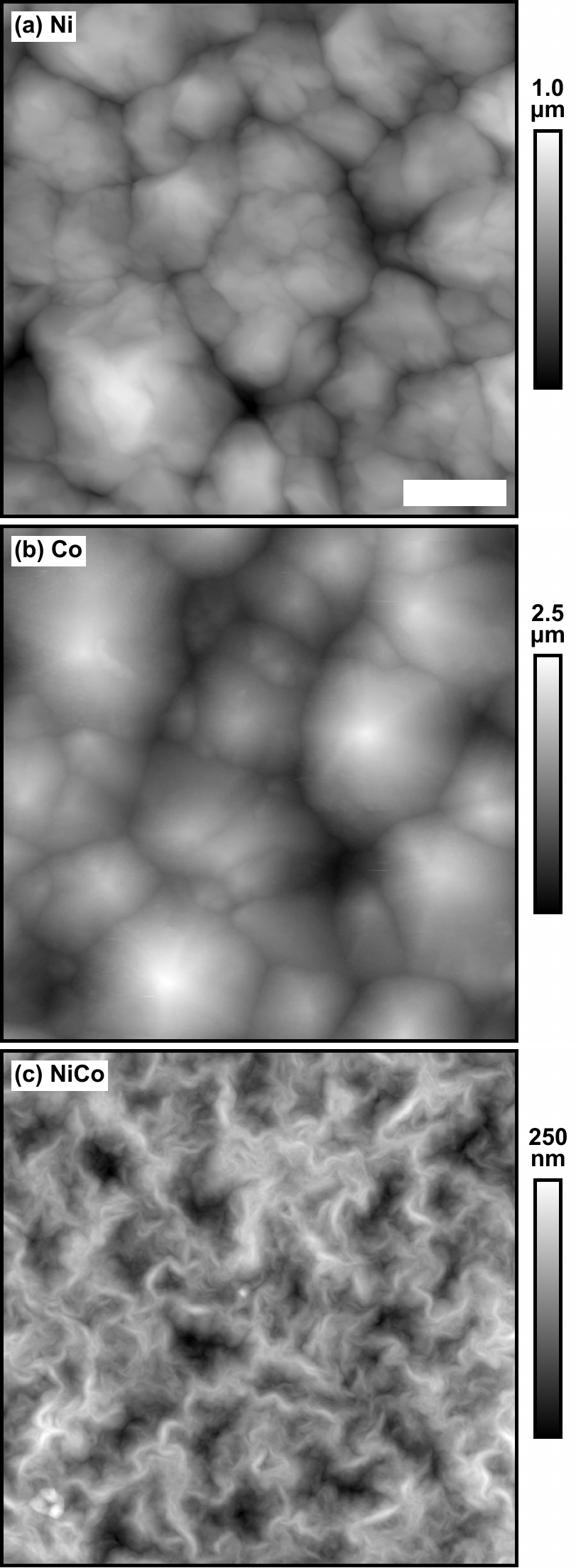}
    \caption{\label{fig:AFMEx} \csentence{Example 10~$\mu$m $\times$ 10~$\mu$m AFM topographic measurements for (a) Ni (b) Co and (c) NiCo thin films.} Each sample had deposited charge of 1000~mC. The scale bar is 2~$\mu$m for all the images. The vertical scale is indicated to the right and is different for each image. The roughness factors for these images are (a) 1.12, (b) 1.41, and (c) 1.05.}
\end{figure}

\begin{figure}[h!]
    \includegraphics{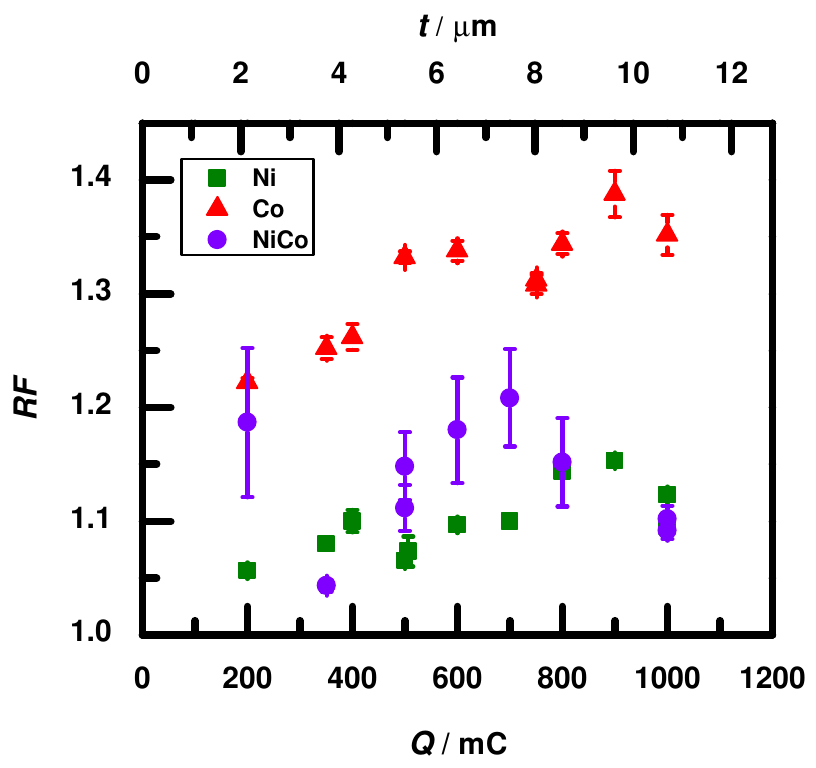}
    \caption{\label{fig:RFvQ} \csentence{Average roughness factor, $RF$, of each sample as a function of the deposited charge, $Q$.} The second horizontal axis indicates the approximate average thickness, $t$, of the samples. Error bars represent the standard error of the mean for the measurements.}
\end{figure}

\begin{figure}[h!]
    \includegraphics{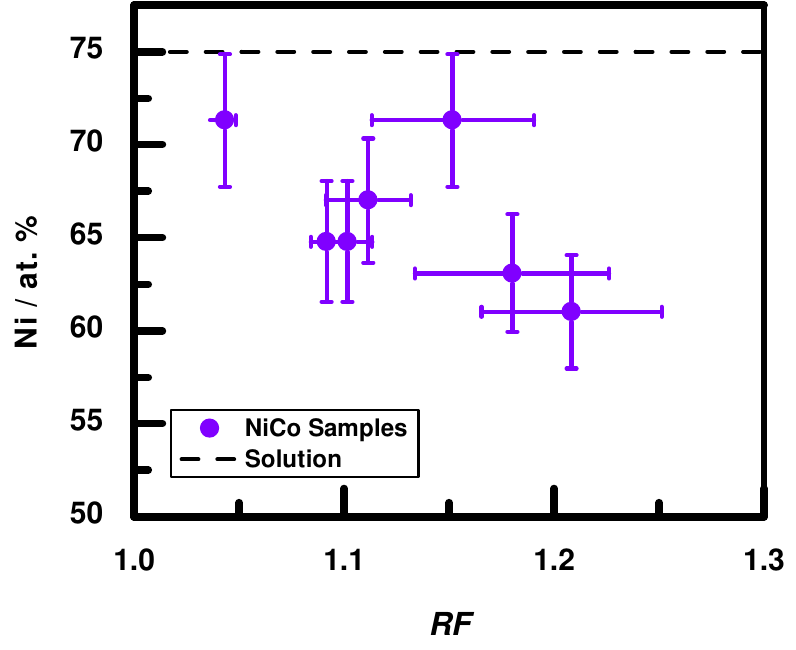}
    \caption{\label{fig:NivsRF} \csentence{Ni composition for the NiCo samples as a function of the average roughness factor, $RF$, of the samples.} Composition error bars represent the typical EDS uncertainty. The dashed line indicates the Ni composition in the deposition solution.}
\end{figure}

\begin{figure}[h!]
    \includegraphics{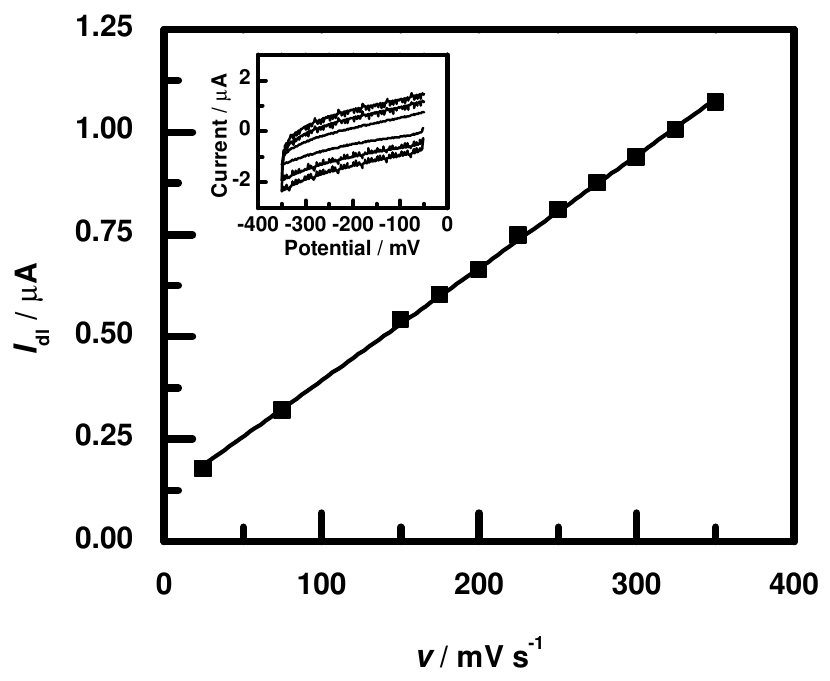}
    \caption{\label{fig:CapEx} \csentence{Example double-layer capacitance measurements for a NiCo thin film.} The sample had a deposited charge of 1000~mC. The inset shows CV measurements in 1~M \ce{KOH}  at 75, 225, and 350~mV/s. The slope of the linear fit to the capacitive current, $I_\mathrm{dl}$, vs.\ scan rate, $v$, is the measured double-layer capacitance, $C_\mathrm{dl}$ , for the sample.}
\end{figure}

\begin{figure}[h!]
    \includegraphics{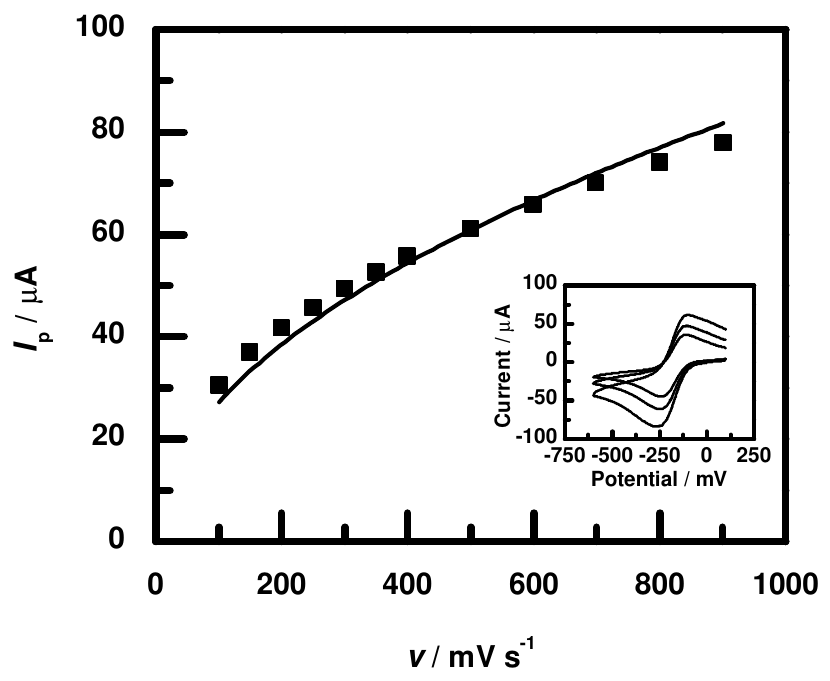}
    \caption{\label{fig:AreaEx} \csentence{Example electrochemical area measurements for a NiCo thin film.} The sample had a deposited charge of 1000~mC. The inset shows CV measurements in 5~mM \ce{Ru(NH3)6Cl3}  and 1~M \ce{KCl} at 200, 400, and 800~mV/s. The magnitude of the peak cathodic current, $I_\mathrm{p}$, is fit to a square root function vs.\ scan rate, $v$, to determine the area, $A_\mathrm{ec}$, of the sample using the Randles-Sevcik equation.}
\end{figure}

\begin{figure}[h!]
    \includegraphics{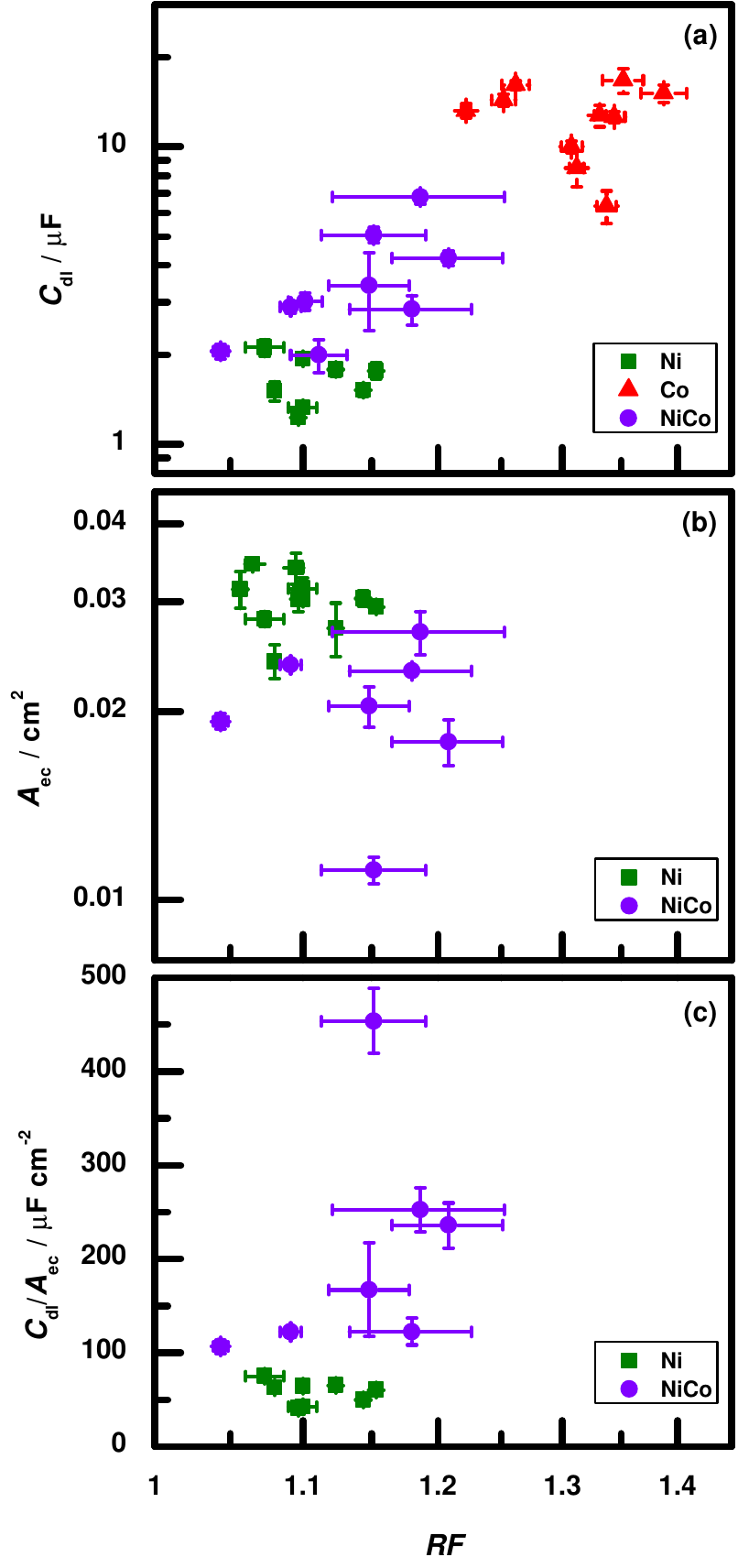}
    \caption{\label{fig:CAvsRF} \csentence{(a) Average capacitance, $C_\mathrm{dl}$, (b) average area, $A_\mathrm{ec}$, and (c) ratio of capacitance to area, $C_\mathrm{dl}/A_\mathrm{ec}$, of each sample as a function of the average roughness factor, $RF$.} Error bars represent the standard error of the mean for the measurements.}
\end{figure}

\end{backmatter}

\end{document}